\documentclass[twocolumn,prl,showpacs]{revtex4}

\usepackage{graphicx}

\begin{document}

\title{Interaction of spatial solitons  in nonlinear optical medium}

\author{R.\ Khomeriki}
\thanks{khomeriki@hotmail.com}

\affiliation{Department of Physics, Tbilisi
State University, Chavchavadze ave.\ 3, Tbilisi 380028, GEORGIA}

\author{L.\ Tkeshelashvili}
\thanks{lasha@tkm.physik.uni-karlsruhe.de}

\affiliation{Institut f\"ur Theorie der Kondensierten Materie,
University of Karlsruhe, P.O. Box 6980, 76128 Karlsruhe, GERMANY;
\\ Institute of Physics, Tamarashvili str.\ 6, Tbilisi 380077, GEORGIA}

\begin{abstract}
\noindent The effects caused by nonresonant nonlinear  interaction between noncollinear self-focusing beams are considered in 2D  optical 
samples using multi-scale analysis.  The analytical expression for the  beams trajectories shift due to the  mutual interaction is  derived, and the range of parameters is given  beyond which the mentioned consideration fails. We compare our results with  the  naive  geometrical optics model. It is shown that these two approaches give the same results. This  justifies  the use  of   geometrical optics approach for description of   elastic and almost-elastic collision processes both in Kerr and saturable nonlinear  media.

\end{abstract}

\vspace{5mm}

\pacs{42.65.Tg, 42.65.-k, 05.45.Yv}

\maketitle

Nonlinear localized waves, or soliton-like excitations play important role  in many  branches of  physics: nonlinear optics \cite{agrawal},  plasma physics \cite{dodd}, hydrodynamics \cite{kuznetsov}, magnetic systems \cite{demokritov}  etc. In contrast with  linear excitations  such nonlinear creations may be  exceedingly stable. That is, they can propagate over long distances without distortion. However most exciting feature of soliton phenomena is their interaction processes. In particular,  when they collide with other ones, solitons  exhibit partical-like behavior \cite{stegeman}. 

Despite the fact that there is a large diversity of nonlinear  physical systems exhibiting soliton like excitations, due to universal properties of such creations,  nonlinear localization dynamics can be described only within a few theoretical  models \cite{dodd}. This fact is of  great importance. In particular, this allows one to study experimentally nonlinear phenomena in most convenient physical systems, while the direct experimental  investigation of the particular system the subject of interest may  be more difficult or even  impossible. To date, such "model" experimental systems are nonlinear spin waves in ferromagnetic films \cite{demokritov}  and spatial  optical solitons \cite{stegeman}. However, magnetic envelope solitons can be observed only in (quasi) one dimensional samples and due to transverse instabilities they are unstable in  higher space  dimensions \cite{demokritov}. Thus  recently suggested interaction  effects  \cite{ramaz}   of noncollinearly propagating 1D envelope solitons in 2D magnetically ordered samples doubtfully will be easily realized experimentally. On the other hand nonlinear  optical medium is most appropriate for the 
mentioned purpose (study 1D solitons noncollinear interaction in 2D samples). In particular, spatial optical solitons exhibit richness of  characteristics  not found for localized  waves  in other nonlinear media \cite{stegeman}.  Indeed, much theoretical and experimental investigations have been performed  for optical  spatial solitons:  elastic interaction of Kerr solitons \cite{zakharov,zakharov1,reynaud,aitchison1}, almost elastic and inelastic  collisions of solitons in saturable media including fusion, fission, annihilation, and spiraling occurances \cite{shih} (see also \cite{stegeman}).

In the present paper we consider theoretically  the problem of nonresonant interaction of Kerr spatial optical solitons. The method used here has been suggested for analytical description of interaction of noncollinearly propagating 1D 
envelope solitons of magnetization in 2D magnetically ordered samples \cite{ramaz}. This 
method itself is a generalization of well-known 1D multiple scale 
perturbation theory \cite{taniuti,yajima} for higher space dimensions. Since  this approach  allows us to study the case of interaction of  two spatial solitons with different carrier wave frequencies, for the  nonresonant interaction of solitons  the results presented here  are  more general compared to ones  obtained in \cite{zakharov}, where the exact solutions are found but they concern only the case of spatial solitons interaction with the same carrier wave number. Later these analytical results has been used for suggesting different applications of spatial soliton interactions, e.g. producing nonlinear photonic switching \cite{shi} and all optical logic elements \cite{tech}. Optical solitons dragging and collisions in the presence of absorptions has been also studied \cite{blair} applying numerical methods. In the present paper we obtain analytical results for collisions of spatial solitons with different carrier wave number and point out possible relevance of this study to the above-mentioned applications.

For simplicity we consider the interaction  of spatial  optical solitons in isotropic  thin optical films and assume that the  electric field is normal to the film plane. If the medium is off-resonant with 
respect to the optical signal and the optical film is thin enough, then the dispersion can be neglected  and modulations develop only along (single) transverse direction \cite{reynaud,aitchison1}. As a result, in the case of appropriate sign of the  nonlinear coefficient,  so called 1D spatial solitons (self-focusing beams) are  formed  in 2D samples. 
Obviously one can consider the crossing of two beams and study analytically 
the influence of one self-focusing beam on the other one  using the above 
mentioned method \cite{ramaz}. 

\begin{figure}[b]
\begin{center}\leavevmode
\includegraphics[width=0.8\linewidth]{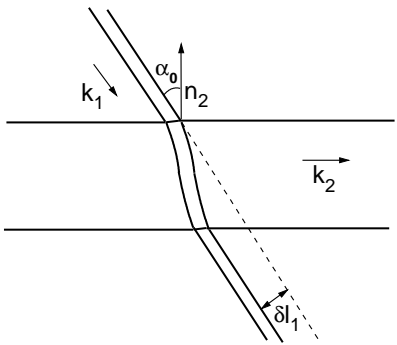}
\vspace{-4.5cm}\noindent
\caption{Schematic   picture of  the interaction process between 
self-focusing beams in off-resonant optical medium. Solid lines indicate 
"borders" of the beams. $\alpha_0$ defines an angle between first (narrow) 
beam and normal vector $n_2$ of the second (wide) beam; $\vec k_1$ and $\vec 
k_2$ are their carrier wave vectors, respectively; $\delta l_1$ stands for a 
shift of trajectory of the first beam, which is caused by the nonlinear 
interaction effects. Let us mention that the second beam trajectory is 
also slightly shifted.} \end{center}
\label{beams1}
\end{figure}

One could try to understand  the  nonlinear effect of spatial solitons  interaction  from the naive geometrical optics model: the propagation of the intense  beam through the sample locally causes the increasing of the refraction index due to the  nonlinear reaction of the medium (Kerr effect).  This in turn  generates wave-guiding area and as a consequence the  beam becomes self-focusing \cite{stegeman}. At the same time, as long as the refraction index within the beam area is bigger than outside it is natural to expect that the second beam will be bent during   crossing the first beam area and eventually  its trajectory will be shifted after the interaction  as  is shown schematically  in Fig. 1.  From the same geometrical optics consideration  it follows that this shift should be zero for perpendicular beams. However, the interaction process is much  more complicated. Actually the second beam affects the induced waveguide of the first beam (the first (wide)  beam in Fig. 1 is slightly  shifted as well).  This gives rise to "self-action" of the beam through other one during the interaction. In addition, interference effects may  take place  between the carrier waves of the interacting beams.  For Kerr spatial solitons it is well  established that for large enough converging  input angles the solitons pass through each other unaffected.  The only effect of such nonresonant interaction is the trajectory  shift of the interacting beams \cite{stegeman}. Thus, qualitatively, the effect is the same as it comes out from the  naive geometrical optics consideration. Surprisingly, in this 
paper we find that the results are the same even quantitatively.  

In nonlinear Kerr  medium polarization $P$ depends nonlinearly on 
electric field $E$ as follows:
\begin{equation}
P=\chi^{(1)}E+\chi^{(3)}E^3, \label{0}
\end{equation}
where $\chi^{(1)}$ and $\chi^{(3)}$ are linear and nonlinear polarisability
constants respectively. For simplicity we consider here centrally symmetric materials and therefore from the symmetry reasons the second order term in Exp. (\ref{0}) is identically zero.
The  wave equation for the  electric field reads (see for more details 
e.g.  \cite{dodd}):
\begin{equation}
\nabla^2E-\beta E_{tt}=\gamma(E^3)_{tt} \label{1}
\end{equation}
with coefficients $\beta=(1+4\pi\chi^{(1)})/c^2$ and 
$\gamma=4\pi\chi^{(3)}/c^2$. The nabla operator acts in 2D space as long as 
film samples are considered in the present paper.

Let us consider the weakly nonlinear case i.e. when the cubic term is much smaller than the linear one. We do not repeat all the steps of calculations to obtain the 1D spatial soliton solution, let us just mention that a weakly nonlinear wave  with a slowly 
varying envelope is sought  in the following way \cite{taniuti}:
\begin{equation}
E=\sum\limits_{\alpha=1}^{\infty} \varepsilon^{\alpha}
\sum\limits_{l=-\infty}^{\infty}
\varphi^{(\alpha)}_l \bigl(\vec\xi,\vec\tau \bigr)
\cdot e^{il(\vec k\vec r - \omega t)}, \label{2}
\end{equation}
where frequency $\omega$ and wave vector $\vec k$ of carrier wave are 
connected via simple dispersion relation $\omega=k/\sqrt{\beta}$; the 
envelope $\varphi^{(\alpha)}_l=\left(\varphi^{(\alpha)}_{-l}\right)^*$ is a 
function of slow variables $\vec\xi=\varepsilon( \vec r-\vec v t)$ and 
$\vec\tau=\varepsilon^2\vec r/2k$; $\vec v=d\omega/d\vec k$ is a group 
velocity of linear wave $\vec v\parallel \vec k$ and $\varepsilon$ is a 
formal parameter defining the smallness or "slowness" of the physical 
quantity before which it appears. Then building the ordinary perturbation 
scenario \cite{taniuti,dodd} in the third approximation over $\varepsilon$ one gets 
1D nonlinear Schr\"odinger (NLS) equation:
\begin{equation}
i \frac{\partial \varphi^{(1)}_1}{\partial \tau} +
\frac{\partial^2 \varphi^{(1)}_1}{\partial \xi^2} + 
3\gamma\omega^2\varphi^{(1)}_1
{\left|\varphi^{(1)}_1\right|}^2 =0, \label{3}
\end{equation}
which has well-known spatial soliton (self-focusing beam) solution if $\gamma>0$. Physically, the spatial soliton formation is the result of balance between defocusing  diffraction and focusing  nonlinearity. It is worth to note that, since diffraction in general is  strong, the nonlinearities involved in spatial solitons  are much  stronger  compared to nonlinearities involved in temporal solitons.   As we see from (\ref{3})  the soliton envelope in the leading approximation is a function of variables $\xi\equiv\vec\xi_\perp=\varepsilon(\vec n \vec r)$ and $\tau\equiv\vec\tau_\parallel=\varepsilon^2(\vec k\vec r)/2k^2$ where $\tau$ plays a role of "time" and $\vec n$ is a unit vector perpendicular to $\vec k$. Let us emphasize once again that we have got 1D NLS due to fact that the longitudinal dispersion in off-resonant optical medium could be neglected, 
in other words wave group velocity $v\equiv d\omega/dk$ does not depend on 
wave number $k$. In case of presence of longitudinal dispersion the physical 
process would be described \cite{dodd} by 2D NLS for which 1D solitonic 
solution would be unstable.

Now we shall start a main task of the paper, particularly, the analytical 
investigation of interaction between noncollinear self-focusing beams. For 
that purpose we are seeking for the solution following the general method 
developed in Refs. \cite{ramaz}:
\begin{equation}
E=\sum\limits_{\alpha=1}^{\infty} \varepsilon^{\alpha}
\sum\limits_{l_1l_2=-\infty}^{\infty}
\varphi_{l_1 l_2}^{(\alpha)} e^{i\left(
{\vec k}_{l_1l_2}{\vec r} - \omega_{l_1l_2}t+
\varepsilon\Omega_{l_1l_2} \right)}, \label{4}
\end{equation}
where $\omega_{l_1l_2}=l_1\omega_1+l_2\omega_2$; ${\vec k}_{l_1l_2}=l_1{\vec 
k}_1+l_2{\vec k}_2$; $\omega_1$, $\vec k_1$ and $\omega_2$, $\vec k_2$ are 
carrier frequencies and wave vectors, respectively; $\varphi_{l_1 
l_2}^{(\alpha)}$ and $\Omega_{l_1l_2}$ are functions of slow variables (p=1, 
2)
\begin{equation}
\xi_p=\varepsilon \Bigl[(\vec n_p\vec r)-
\varepsilon\psi_p\bigl(\xi_1,\xi_2,
\tau_1,\tau_2\bigr) \Bigr], \quad \tau_p=\varepsilon^2\frac{(\vec k_p\vec 
r)}{2k_p^2}, \label{5}
\end{equation}
where $\vec n_1$ and $\vec n_2$ are unit vectors perpendicular to carrier 
wave vectors $\vec k_1$ and $\vec k_2$, respectively. Proceeding further 
with the similar calculations as it was done in Refs. \cite{ramaz} we come 
in the leading approximation to the following solution:
\begin{equation}
E=\varphi_{10}^{(1)} e^{i\left(
{\vec k}_{1}{\vec r} - \omega_{1}t+
\Omega_{10}^{(1)} \right)}+ \varphi_{01}^{(1)} e^{i\left(
{\vec k}_{2}{\vec r} - \omega_{2}t+
\Omega_{01}^{(1)} \right)}+c.c., \label{6}
\end{equation}
where "c.c" denotes complex conjugated terms; $\varphi_{10}^{(1)}$ and 
$\varphi_{01}^{(1)}$ are the solutions of 1D NLS (see Eq. (\ref{3})) with 
derivatives over set of slow variables $\xi_1$, $\tau_1$ and $\xi_2$, 
$\tau_2$, respectively. For example, in the simplest case of one-soliton 
envelopes $\varphi_{10}^{(1)}$ and $\varphi_{01}^{(1)}$ we have:
\begin{eqnarray}
|\varphi_{10}^{(1)}|&=& |A_1|\cdot
sech \Bigl\{|A_1|\sqrt{6\gamma\omega_1^2}\bigl[(\vec n_1\vec 
r)-\psi_1^{(1)}\bigr]\Bigr\}, \nonumber \\ 
|\varphi_{01}^{(1)}|&=& |A_2|\cdot
sech \Bigl\{|A_2|\sqrt{6\gamma\omega_2^2}\bigl[(\vec n_2\vec 
r)-\psi_2^{(1)}\bigr]\Bigr\},
\label{7}
\end{eqnarray}
and $A_1$ and $A_2$ are amplitudes of the  first and the  second self-focusing beams, respectively. Besides that, phase and position  shifts of the first self focusing beam induced by  weakly nonlinear interaction with the second 
beam are defined by the following formulas:
\begin{equation}
\frac{\partial\psi_1^{(1)}}{\partial\xi_2}=\frac{(\vec n_1\vec n_2)}{(\vec 
k_1\vec n_2)} 
\frac{\partial\Omega_{10}^{(1)}}{\partial\xi_2}=3\gamma\omega_1^2|\varphi_{01}^{(1)}|^2\frac{(\vec 
n_1\vec n_2)}{(\vec k_1\vec n_2)^2}. \label{8}
\end{equation}
As far as according to perturbative approach we have a following scaling 
$\partial\psi_1^{(1)}/\partial\xi_2\sim \varepsilon^2$ and taking into 
consideration the dispersion relation ($\omega_1^2=k_1^2/\beta$) the 
following restriction upon the soliton parameters is derived:
\begin{equation}
|A_2|\sqrt{\frac{3\gamma}{\beta}\frac{|\sin\alpha_0|}{\cos^2\alpha_0}}\ll 1, \label{9}
\end{equation}
where $\alpha_0=(\pi/2)-\theta_0$, and  $\theta_0$ is an angle between the self-focusing beams.  From Exp. (\ref{8}) it is easy to get analytical expression for trajectory shift of 
the first beam caused by the  nonresonant interaction with other one:
\begin{eqnarray}
\delta l_1&=&\psi_1^{(1)}(\infty)-\psi_1^{(1)}(-\infty) \nonumber \\ 
&=&\frac{3\gamma}{\beta}\frac{\sin\alpha_0}{\cos^2\alpha_0}\int\limits_{-\infty}^\infty|\varphi_{01}^{(1)}(\xi_2)|^2d\xi_2. 
\label{10}
\end{eqnarray}

\begin{figure}[t]
\begin{center}\leavevmode
\includegraphics[width=0.8\linewidth]{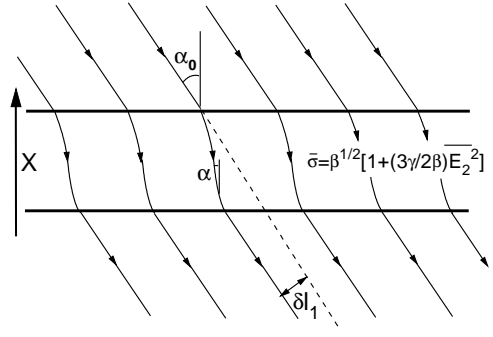}
\vspace{-4.5cm}\noindent
\caption{Optical ray refraction through the self-focusing beam area with 
borders denoted by horizontal solid lines.} \end{center}
\label{beams2}
\end{figure}

Particularly, in case of one-solitonic envelopes (\ref{7}) one gets from (\ref{10}) the analytical expression for the trajectory shift of the first spatial soliton:
\begin{equation}
\delta l_1=|A_2|\frac{\sqrt{6\gamma}}{\beta\omega_2}\frac{\sin\alpha_0}{\cos^2\alpha_0}
\label{anal}
\end{equation}
In addition we want to emphasize that   beyond the limit given by  (\ref{9})  the dynamics is governed by a  set of  two coupled NLS type equations, which in general is not integrable model and gives qualitatively different behavior of interacting solitons (see \cite{cohen} and discussion there).

Now let us compare the results obtained above with  the picture given by  the naive geometric optics consideration. This will  give us  more deep insight to the problem.  First  suppose that one has only one 
self-focusing beam (particularly, the second (wide) beam) and let us calculate how it changes refraction index (see Fig. 2). In view of dependence of 
polarization upon the electric field (\ref{0}) we can write down the 
expression for refraction index as $\sigma=\sqrt{\beta+3\gamma E_2^2}$, 
where $E_2$ denotes electric field in the self-focusing beam area and it has 
a form of envelope spatial soliton (\ref{7}). Averaging refraction index 
over fast variables $\vec r$ and $t$ in the  weakly nonlinear limit (the term 
proportional to $E_2^2$ is small) we get the following approximate formula 
for slowly varying (along axis $x$) averaged refraction index 
$\bar\sigma(x)=\sqrt{\beta}\bigl[1+(3\gamma/2\beta)\overline{E_2^2}\bigr]$. 

Let us now  solve the following  geometrical optics problem, particularly, how the optical rays refract propagating through the area of the second beam  with slowly changing refraction index. For that purpose we  note that an angle $\alpha$ between the ray 
and normal vector (with respect to the beam) at  any point could be 
calculated via simple refraction formula: 
$\sigma(x)\sin\alpha=\sigma_0\sin\alpha_0$, where 
$\sigma_0\equiv\sqrt{\beta}$. Taking into account the fact that the trajectory shift of the ray could be calculated as follows: $\delta 
l_1=\cos\alpha_0\int\limits_{-\infty}^\infty dx(\tan\alpha_0-\tan\alpha)$, we 
come exactly to the formula (\ref{10}) which we have obtained from multi-scale analysis \cite{ramaz}. However, to avoid any misunderstanding  it has to be stressed  that  the geometrical optics  approach to the  spatial solitons interaction problem  is not self-consistent. Indeed, as we have pointed out  already,   in this model  the "self-action" effects of the first soliton  through other one are  neglected.  Physically this  means  that  the first beam is linear. But the diffraction in the nonlinear   problem considered here in not negligible. Thus the first beam will diffract and the interaction picture given by the  geometrical optics model is not meaningful under realistic experimental situations of the  spatial solitons interaction. 

  Although the geometrical optics approach  in some particular cases gives full understanding  of the   solitons  interaction processes \cite{snyder}, in general one expects that this approach is valid only for  qualitative description of the incoherent  spatial  solitons interaction \cite{stegeman}. As is  mentioned above the  nonlinear "self-action" of the beam through other one  is neglected without justification in the  geometrical optics model of solitons collisions. However, the analysis presented here shows that this additional  nonlinear "self-action" during the  interaction process does not affect the  soliton dynamics asymptotically. This is why such naive geometrical optics model gives correct results even for almost-elastic collision processes  between solitons in saturable media \cite{stegeman}.

In summary, in the present paper we have considered  the problem of nonresonant interaction of Kerr spatial solitons (self-focusing beams)  with different carrier wave frequencies  using multi-scale analysis. It is shown that the beams trajectories are shifted due to mutual interaction. The analytical expressions for these shifts are obtained as well.  Surprisingly the naive geometrical optics model of the solitons collision is in full agreement with  the results of general theory. In particular,  this  shows that the "self-action" of the  soliton caused by  nonresonant interaction process does not changes it's  asymptotical behavior after the collision. This in turn justifies  the use  of the  geometrical optics model  for description of elastic and almost-elastic collision processes both in Kerr and saturable media.

R. Kh. is obliged to Greg Salamo for the enlightening discussions concerning the physics of self-focusing beams interaction process. L.T.  acknowledges financial support from the Deutsche Forschungsgemeinschaft (DFG)  under Bu 1107/2-1 (Emmy-Noether program) and the DFG-Forschungszentrum Center for Functional Nanostructures (CFN) at the University of Karlsruhe. R. Kh. is supported by USA Civilian Research and Development Foundation award No GP2-2311-TB-02.


\begin{thebibliography}{99}
\bibitem{agrawal} G. P. Agrawal, {\it  Nonlinear Fiber Optics}, 2nd ed. (Academic Press, San Diego, 1995).
\bibitem{dodd} R. K. Dodd, J. C. Eilbeck, J. D. Gibbon, and H. C. Morris, {\it Solitons and Nonlinear Wave Equations} (Academic Press, London, 1982).
\bibitem{kuznetsov} V. E. Zakharov and E. A. Kuznetsov, Physics-Uspekhi  {\bf40}, 1087 (1997).
\bibitem{demokritov} S. O. Demokritov, B. Hillebrands, and A. N. Slavin, Phys. Rep. {\bf348}, 441 (2001).
\bibitem{stegeman} G. I. Stegeman and M. Segev, Science {\bf286}, 1518 (1999). 
\bibitem{ramaz} N.  Giorgadze and  R. Khomeriki, J. Magn. Magn. Mater.{\bf 186}, 239 (1998); N. Giorgadze and  R. Khomeriki, Phys. Rev. B. {\bf
60}, 1247 (1999); R. Khomeriki and  L. Tkeshelashvili, J. Phys.: Condens. 
Matter {\bf 12}, 8875 (2000).
\bibitem{zakharov} V. E. Zakharov and A. B. Shabat, Sov. Phys.  JETP {\bf34}, 62 (1972). 
\bibitem{zakharov1} V. I. Karpman  and V. V. Solov'ev, Physica D  {\bf3}, 487 (1981); D. Anderson  and M. Lisak, Phys. Rev. A  {\bf32}, 2270 (1985); 
\bibitem{reynaud} F. Reynaud and A. Barthelemy, Europhys. Lett. {\bf12}, 401 (1990); 
\bibitem{aitchison1} J. S. Aitchison et al., Opt. Lett. {\bf16}, 15 (1991); 
J. Opt. Soc. Am. B {\bf8}, 1290 (1991).
\bibitem{shih} M. Shih and M. Segev, Opt. Lett. {\bf21}, 1538 (1996); V. Tikhonenko, J. Christou, and B. Luther-Davies, Phys. Rev. Lett. {\bf76}, 2698 (1996); H. Meng et al., Opt. Lett. {\bf22}, 448 (1997); W. Krolikowski and S. A. Holmstrom, Opt. Lett. {\bf22}, 369 (1997); M. Shih, M. Segev, and G. Salamo, Phys. Rev. Lett. {\bf78}, 2551 (1997); A. Buryak et al., Phys. Rev. Lett. {\bf82}, 81 (1999).
\bibitem{taniuti} T. Taniuti, Progr. Theor. Phys. (Suppl.) {\bf 55}, 1  
(1974). 
\bibitem{yajima} M. Oikawa and  N. Yajima, J. Phys. Soc. Japan  {\bf 37}, 486  (1974).
\bibitem{shi} T.-T. Shi, S. Chi, Opt. Lett., {\bf 15}, 1123, (1990).
\bibitem{tech} O. V. Kolokoltsev, R. Salas, and V. Vountesmeri,  J. Lightwave Technology, {\bf  20}, 1048, (2002).
\bibitem{blair} S. Blair, K. Wagner R. McLeod, J. Opt. Soc. Am. B, {\bf 13}, 2141, (1996).
\bibitem{cohen} O. Cohen et al., Phys. Rev. Lett. {\bf 89}, 133901 (2002).
\bibitem{snyder} A. W. Snyder and D. J. Mitchell, Science {\bf 276}, 1538 (1997).



\end{thebibliography}
\end{document}